# Strain engineering in Ge/GeSn core/shell nanowires

S. Assali,[1,2,¥,*] M. Albani,[3,¥,*] R. Bergamaschini,[3] M. A. Verheijen,[1,4] A. Li,[1,5,6]

S. Kölling,[1] L. Gagliano,[1] E. P.A.M. Bakkers,[1,6] and L. Miglio[3]

[1] Department of Applied Physics, Eindhoven University of Technology, 5600 MB Eindhoven, The Netherlands
[2] Department of Engineering Physics, École Polytechnique de Montréal, C. P. 6079, Succ. Centre-Ville, Montréal, Québec H3C 3A7, Canada
[3] L-NESS and Dept. of Materials Science, University of Milano Bicocca, 20125, Milano, Italy
[4] Eurofins Materials Science BV, High Tech Campus 11, 5656AE Eindhoven, The Netherlands
[5] Beijing University of Technology, Pingleyuan 100, 100124, P. R. China
[6] Kavli Institute of Nanoscience, Delft University of Technology, 2600 GA Delft, The Netherlands

Strain engineering in Sn-rich group IV semiconductors is a key enabling factor to exploit the direct band gap at mid-infrared wavelengths. Here, we investigate the effect of strain on the growth of GeSn alloys in a Ge/GeSn core/shell nanowire geometry. Incorporation of Sn content in the 10-20 at.% range is achieved with Ge core diameters ranging from 50nm to 100nm. While the smaller cores lead to the formation of a regular and homogeneous GeSn shell, larger cores lead to the formation of multi-faceted sidewalls and broadened segregation domains, inducing the nucleation of defects. This behavior is rationalized in terms of the different residual strain, as obtained by realistic finite element method simulations. The extended analysis of the strain relaxation as a function of core and shell sizes, in comparison with the conventional planar geometry, provides a deeper understanding of the role of strain in the epitaxy of metastable GeSn semiconductors.

Strained semiconductor heterostructures provide a rich playground for investigating the epitaxy of lattice-mismatched materials.[1] In the last decades SiGe alloys grown with a graded composition on Si were extensively studied to relieve strain by nucleating misfit dislocations in the buffer layers.[2–4] Recently, direct band gap and metastable GeSn alloys gained tremendous interest as a platform for Si-compatible photonics operating at mid-infrared wavelengths.[5–9] In unstrained GeSn the direct band gap is achieved at Sn contents higher than 10at.%, hence well above the ~1at.% equilibrium solubility of Sn in Ge. The incorporation of Sn is highly sensitive to the *in-plane* strain that the GeSn layer experiences during growth.[10,11] Due to the large lattice mismatch between Ge and α-Sn (>10%), the growth of GeSn layers has been developed on high-quality Ge-virtual substrates (Ge-VS) on Si.[12,13] Partial strain relaxation can induce a compositional grading in GeSn,[8,14–16] eventually leading to segregation and precipitation of Sn, compromising material quality.[17–19] In GeSn layers grown on Ge-VS, the compressive strain is reduced in a multi-layer buffered heterostructure grown with different Sn contents by controlling the growth temperature[20,21] and precursors flows.[22] The high amount of strain induces nucleation of dislocations in the low (7-11at.%) Sn content buffer layers,[11,23] and the resulting uniform (plastic) strain relaxation enhances the Sn incorporation above 16at.% in the GeSn layers grown on top.[8,11,14,19] One-dimensional nanowires(NWs) provide additional degrees of freedom in tuning the effect of strain in the growth of lattice-mismatch heterostructures[24,25] when using a core/shell NW geometry.[26] The shell displays an increasing strain relaxation with thickness provided by the free surfaces at the sidewalls, while the elastic compliance of the NW core allows for enhanced strain relaxation in the shell, accommodating the lattice mismatch of the system and avoiding bending.[26,27] Recent studies on Ge/GeSn core/shell NWs[15,16,28,29] are mainly focused on small Ge core-sizes, where a low amount of residual strain is induced in the GeSn shell.

In this Letter, we show how strain can be engineered by tuning the core and shell sizes and we explore high strain conditions focusing on large cores and high Sn contents. To this purpose, core diameters ranging from 50nm to 100nm are considered for the growth of the GeSn shell and the samples are analyzed using transmission electron microscopy(TEM) to assess the crystal quality and the Sn incorporation. Realistic Finite Element Method (FEM) simulations are then performed to characterize the strain distribution in the grown samples and to correlate strain partitioning with NW geometry.

The vapor-liquid-solid(VLS) growth of Au-catalyzed arrays of Ge/GeSn core/shell NWs is performed in a chemical vapor deposition (CVD) reactor using germane ($GeH_4$), tin-tetrachloride ($SnCl_4$), and hydrogen chloride (HCl) as precursors (supplementary material S1).[15] The two-temperature step growth is held at 320°C and 300°C for the Ge core and GeSn shell growth, respectively. Three different NW arrays with Ge core diameters of 50nm, 65nm, and 100nm were fabricated by controlling the Au layer thickness during nanoimprint lithography.

The scanning electron microscope (SEM) image in Fig.1a shows an array of $Ge/Ge_{0.895}Sn_{0.105}$ core/shell NWs grown using a 50nm core. Tapered top section and flat {112} sidewalls are obtained, as previously discussed in Refs.[15,16]. An important feature distinguishable in the bottom section of the NWs is a more complex faceting of the shell morphology, which was reported to be more evident for a higher Sn content $Ge_{0.87}Sn_{0.13}$ shell grown at a lower $GeH_4/SnCl_4$ precursor ratio.[15] The extension and morphology of the multi-faceted bottom section is strongly influenced by the diameter of the Ge core, either defined by the size of the Au droplet, or resulting from the small tapering at the NW base. When the Ge core diameter is increased from 50nm to 65nm, the length of the multi-faceted bottom section extends to more than half of the NW length (Fig.1b). With a further increase of the Ge core diameter to 100nm, the shell becomes multi-faceted (Fig.1c) and its thickness decreases to ~60nm. We note that the volume of the GeSn shell grown around the 100nm Ge core is less than one half as compared to when using 50nm cores. In addition, a small degree of tapering is always observed in the Ge core NW arrays, independently of the diameter (supplementary material S2). Therefore, the reduction in shell thickness and volume with the development of a multi-faceted sidewall is associated with the increase in Ge core diameter. This suggests strain-driven growth kinetics during GeSn shell growth, which is a similar situation to what observed in the growth of GeSn in a planar geometry.[8,10,11] Furthermore, the HCl supply during the GeSn shell growth does not contribute to the change in morphology in the bottom section of the NWs (Supplementary material S3).

A detailed insight on the irregular morphology of the GeSn shell grown using 100nm Ge cores is obtained using transmission electron microscopy (TEM). The GeSn shell is visible in the energy-dispersive X-ray (EDX) compositional map acquired in scanning-TEM (STEM) mode in Fig.2a. No axial growth of a GeSn segment is observed,[15] while a GeSn shell with a variable thickness and a multi-faceted sidewall are visible in the TEM images in Fig.2b-d. This faceting cannot be related to the core morphology, as the 100nm Ge-core NWs have flat sidewalls, similar to the 50nm cores (supplementary material S1). Thus, the change in shell morphology between the flat sidewall observed for 50nm cores[15,16] and the multi-faceted sidewall with 100nm cores most likely relates to the larger amount of strain in the shell, as we will quantify in the following. Higher strain beyond a critical value, leads to plastic relaxation, with the nucleation of defects in the core/shell NW heterostructure. Few defects are indeed visible in the core/shell NW in Fig.2b-d, as indicated using red arrows. Due to the large sample thickness a precise identification of the type of defects in the shell, such as partial dislocations, is not possible. It is important to compare this situation with the NWs grown using a 50nm Ge core, where no defect lines were identified in the ~120nm thick GeSn shell.[15] In the growth of the GeSn shell around 100nm Ge cores, multiple defects can be identified in the GeSn shell with a thickness of only ~60nm. Thus, the plastic relaxation in the shell at a smaller thickness and similar Sn content indicates that larger strain energy is present in the core/shell NW during the growth using larger Ge cores. Cross-sectional EDX measurements were performed to determine distribution of Sn across the shell thickness. In the case of 50nm Ge cores the GeSn shell exhibits a hexagonal shape bounded by {112} facets and nm-thin, Sn-poor sunburst stripes along the vertices.[15] On the contrary, as shown in Fig.3, 100nm-core NWs have a more irregular morphology with both {112} and {110} facets. Also the composition becomes inhomogeneous, with a Sn content up to ~21at.% along the radial <112> directions and up to ~10at.% along the radial <110> directions. The precise shape and composition profiles are found to strongly depend on the actual growth conditions, i.e. $SnCl_4$ precursor flow. A detailed understanding of this complex behavior, mostly influenced by the Sn incorporation dynamics, is beyond the scope of the present Letter, and will be addressed in a separate work.[30]

We now focus on the characterization of the residual strain in the GeSn shell. A quantitative determination of the strain distribution in cross-sectional TEM samples using electron diffraction is challenging.[16] Strain imaging in TEM would require both compositional mapping as well as lattice periodicity mapping on the nanometer scale, because of the inhomogeneous Sn incorporation in the GeSn shell. To circumvent this problem, we use a different strategy by estimating the elastic strain relaxation in the core/shell NW system by FEM simulations.[16] The NW is modelled as a dodecagonal Ge core, along the [111] direction, surrounded by the GeSn shell bounded by six main {112} facets and six smaller {110} ones, perpendicular to the substrate (supplementary material S4). The Sn composition is set differently along the <112> and <110> portions as indicated by the EDX maps in Fig.3, including the linear increase with the radius made evident in Fig.3c. The strain in the NW is originating from the (bulk) lattice mismatch between the Ge core, which is the initial template for the epitaxial growth, and the material in the GeSn shell.[31] The NW has three main relaxation mechanisms, according to its symmetry around the axis.

Radially the shell can expand freely toward the free surface, while tangentially it is bounded by a ring geometry around the core, therefore it remains compressed as shown by the color maps in Fig.4a. Along the axial direction, on one side it can expand toward the top free surface, but on the other side it is tied to the axial lattice parameter of the Ge core. As a result, the relaxation of the GeSn shell requires an axial tensile deformation of the core, as it can be seen by the color maps in Fig.4a-b. Simulations have been performed to study the variation during growth of the strain at the surface, *i.e.* where the incorporation of Sn adatoms is active. In particular, in Fig.4c the *in-plane* components (tangential and axial) have been averaged on the surface, for different shell thicknesses. To understand the possible impact of the core diameter on the surface strain different Ge core diameters were compared. A large increase in the *in-plane* compressive strain from -0.3% to -0.9% is observed by changing the Ge core diameter from 20nm to 110nm at a constant GeSn shell thickness of 20nm. Thus, an increase of the overall strain energy in the core/shell system with increasing Ge core diameter is present, which is also maintained at larger shell thicknesses.

Since the Sn composition is not uniform, the evaluation of the residual strain is not simply related to the relative volume of the core and the shell. Therefore, FEM simulations are strictly required for a precise estimation of the NW deformation. The inner part of the shell, having a lower Sn composition, may have even a tensile strain, behaving, in the same way as the core, as a compliant substrate for the outer shell which has a larger volume and a stronger tendency to expand, because of the higher Sn content. This is particularly evident in the color map in Fig.4b for the NW with the smallest Ge core (20nm), which can be easily deformed by the surrounding GeSn shell. The axial strain averaged in the core volume is plotted in Fig.4d as a function of the GeSn shell thickness, for different core diameters. The axial expansion is more than 3x times bigger than the radial and tangential ones, revealing a substantially uniaxial character for the core deformation. It is worth noting that our calculations provide the maximum residual strain within the NW, since purely elastic relaxation is taken into account. However, in the case of larger cores, the strain accumulation in the shell is likely to exceed the onset of plastic relaxation, with the development of extended defects, partially relieving strain (Fig.2c-d). In addition, the critical thickness should be larger than in (not compliant) planar substrates, still the circular periodicity induced by the tubular configuration of the shell is likely to affect the nucleation and multiplication mechanisms of misfit dislocations.[32] This issue would require a detailed HRTEM analysis, not presently viable, as explained above. However, plastic relaxation might easily be responsible for the inhomogeneous shell growth (Fig.2) producing local variations of the surface chemical potential, hence altering the Sn incorporation dynamics.

Lastly, we note that a tensile axial strain above 2% in the 20nm Ge core can be achieved at a GeSn shell thickness larger than 60nm (Fig.4d), which could eventually induce an indirect to direct band gap transition in the Ge core, thus enriching the physical properties of a fully-integrated group-IV semiconductor opto-electronic platform.[33–35]

In conclusion, the growth of GeSn alloys in a Ge/GeSn core/shell NW heterostructure shows few striking differences with respect to conventional planar growth. In the latter, the uniform plastic relaxation allows for enhanced Sn incorporation beyond the dislocated region while keeping a limited surface roughness (<10nm).[8,11,14,19] When using Ge/GeSn core/shell NWs, competing strain relaxation between the (fewer) more geometrically constrained defects and the non-uniform Sn distribution (on the {110}-{112} facets) takes place. Therefore, the growth of GeSn alloys in a core/shell NW geometry is beneficial when the strain in the shell is kept below the threshold for plastic relaxation, hence when using thinner (50nm) Ge cores. On the contrary, for larger (100nm) Ge cores the higher strain in the shell induces a more irregular growth of the GeSn shell, affected by both structural imperfections (multi-faceted and rotated sidewall, defects) and compositional inhomogeneities. These results show the critical role of strain in the growth of GeSn alloys, which can be further investigated with the development of SiGeSn alloys for enhanced strain and band gap engineering.[36,37]

**Figures**

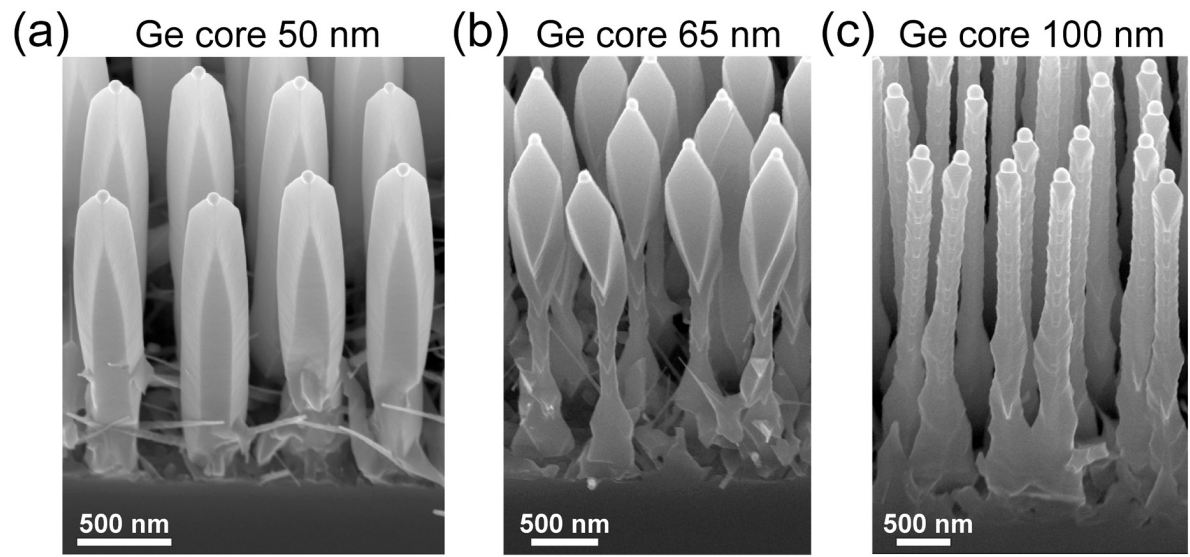

**Figure 1.** (a-c) SEM images of the Ge/GeSn core/shell NW arrays (tilting angle 30°) grown using a Ge core diameter of 50nm(a), 65nm(b), and 100nm(c).

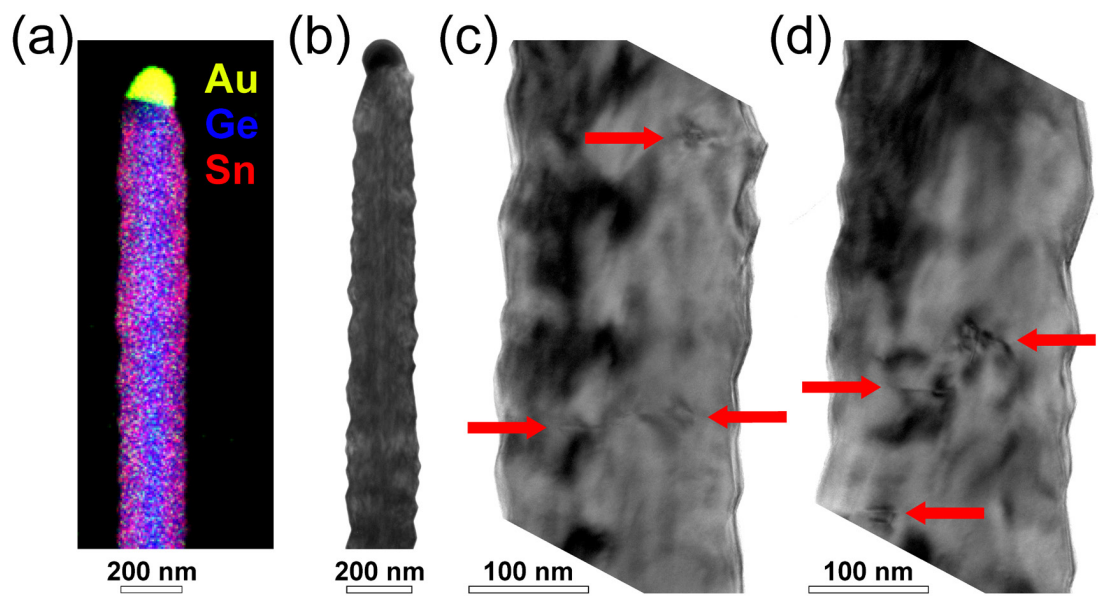

**Figure 2.** (a) EDX compositional map showing the presence of the GeSn shell around the 100nm Ge core. (b-d) Bright-field TEM images acquired along the [110] zone axis of a Ge/GeSn core/shell NW with a 100nm core. The multiple defects are highlighted by red arrows.

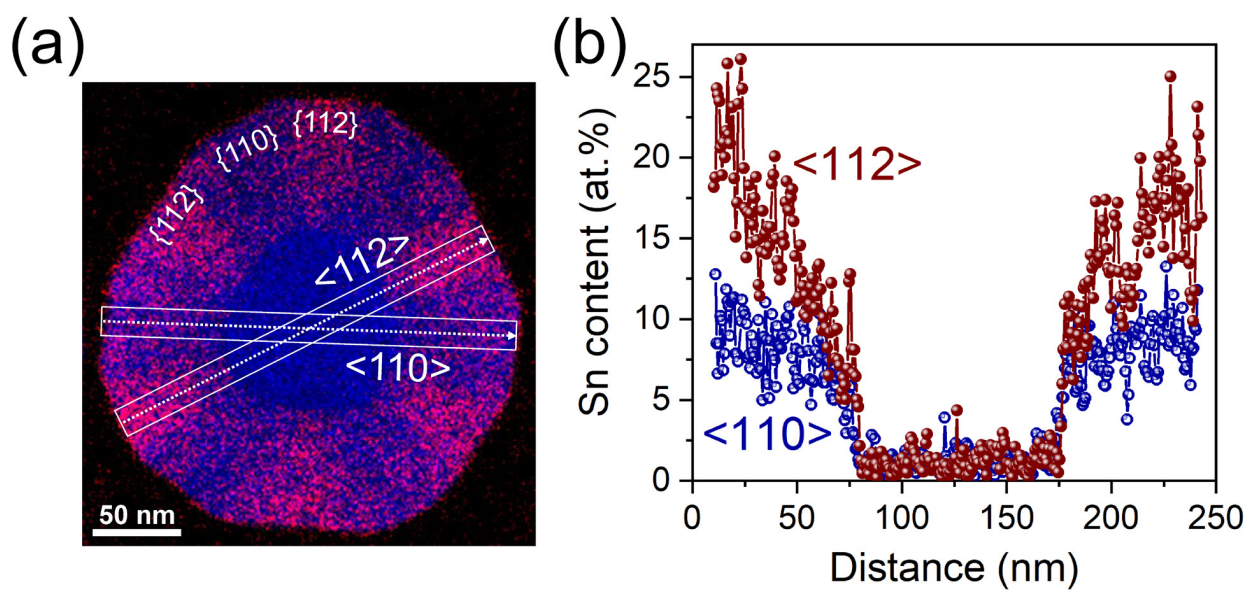

**Figure 3.** (a-b) Cross-sectional EDX compositional map(a) and corresponding plot of the Sn content as a function of the distance along the <112> and <110> radial directions(b).

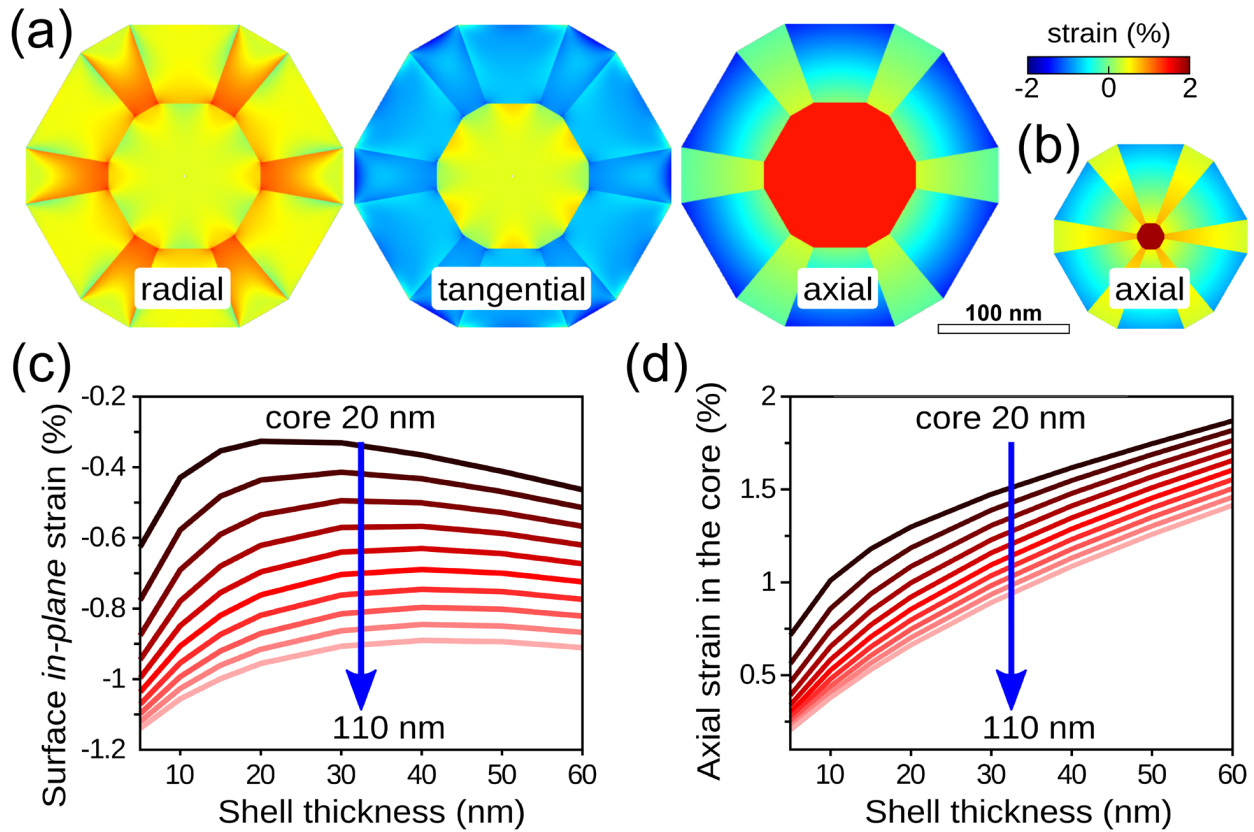

**Figure 4.** (a) Color map of the radial, tangential and axial FEM strain components for a 110nm core diameter and a 60nm-thick shell. (b) Axial strain for a 20/60nm Ge/GeSn core/shell NW heterostructure. (c-d) *In-plane* strain (average of tangential and axial) at the shell surface (c) and axial strain in the Ge core (d) as a function of the shell thickness, for different core diameters. A linear composition gradient along <112> stripes from 9at.% up to 18at.% Sn for a 60nm-thick shell (7at.% to 10at.% Sn along <110> stripes) is assumed to match the experiments.

## Supplementary material

See supplementary material for additional information on the growth conditions, structural characterization, and FEM simulations.

**Acknowledgements**

The authors thank P. J. van Veldhoven for the technical support with the MOVPE reactor. This work was supported by the Dutch Organization for Scientific Research (NWO-VICI 700.10.441), and by the Dutch Technology Foundation (STW 12744) and Philips Research. Solliance and the Dutch province of Noord-Brabant are acknowledged for funding the TEM facility.

**Author information**

[¥] These authors contributed equally to this work.

[*] Corresponding authors: simone.assali@polymtl.ca; marco.albani@unimib.it

Notes:

The authors declare no competing financial interest.

**References**

[1] R.M. France, F. Dimroth, T.J. Grassman, and R.R. King, MRS Bull. **41**, 202 (2016).

[2] S.B. Samavedam and E.A. Fitzgerald, J. Appl. Phys. **81**, 3108 (1997).

[3] M.T. Currie, S.B. Samavedam, T.A. Langdo, C.W. Leitz, and E.A. Fitzgerald, Appl. Phys. Lett. **72**, 1718 (1998).

[4] D.D. Cannon, J. Liu, D.T. Danielson, S. Jongthammanurak, U.U. Enuha, K. Wada, J. Michel, and L.C. Kimerling, Appl. Phys. Lett. **91**, 252111 (2007).

[5] S. Wirths, R. Geiger, N. von den Driesch, G. Mussler, T. Stoica, S. Mantl, Z. Ikonic, M. Luysberg, S. Chiussi, J.M. Hartmann, H. Sigg, J. Faist, D. Buca, and D. Grützmacher, Nat.

Photonics **9**, 88 (2015).

[6] J. Margetis, S. Al-Kabi, W. Du, W. Dou, Y. Zhou, T. Pham, P. Grant, S. Ghetmiri, A. Mosleh, B. Li, J. Liu, G. Sun, R. Soref, J. Tolle, M. Mortazavi, and S.-Q. Yu, ACS Photonics **5**, 827 (2018).

[7] V. Reboud, A. Gassenq, N. Pauc, J. Aubin, L. Milord, Q.M. Thai, M. Bertrand, K. Guilloy, D. Rouchon, J. Rothman, T. Zabel, F. Armand Pilon, H. Sigg, A. Chelnokov, J.M. Hartmann, and V. Calvo, Appl. Phys. Lett. **111**, 092101 (2017).

[8] S. Assali, J. Nicolas, S. Mukherjee, A. Dijkstra, and O. Moutanabbir, Appl. Phys. Lett. **112**, 251903 (2018).

[9] A. Attiaoui, S. Wirth, A.-P. Blanchard-Dionne, M. Meunier, J.M. Hartmann, D. Buca, and O. Moutanabbir, J. Appl. Phys. **123**, 223102 (2018).

[10] J. Margetis, S.-Q. Yu, N. Bhargava, B. Li, W. Du, and J. Tolle, Semicond. Sci. Technol. **32**, 124006 (2017).

[11] S. Assali, J. Nicolas, and O. Moutanabbir, J. Appl. Phys. **125**, 025304 (2019).

[12] J.M. Hartmann, A. Abbadie, J.P. Barnes, J.M. Fédéli, T. Billon, and L. Vivien, J. Cryst. Growth **312**, 532 (2010).

[13] Y. Yamamoto, P. Zaumseil, T. Arguirov, M. Kittler, and B. Tillack, Solid. State. Electron. **60**, 2 (2011).

[14] W. Dou, M. Benamara, A. Mosleh, J. Margetis, P. Grant, Y. Zhou, S. Al-Kabi, W. Du, J. Tolle, B. Li, M. Mortazavi, and S.-Q. Yu, Sci. Rep. **8**, 5640 (2018).

[15] S. Assali, A. Dijkstra, A. Li, S. Koelling, M.A. Verheijen, L. Gagliano, N. von den Driesch, D. Buca, P.M. Koenraad, J.E.M. Haverkort, and E.P.A.M. Bakkers, Nano Lett. **17**, 1538 (2017).

[16] M. Albani, S. Assali, M.A. Verheijen, S. Koelling, R. Bergamaschini, F. Pezzoli, E.P.A.M. Bakkers, and L. Miglio, Nanoscale **10**, 7250 (2018).

[17] J. Aubin, J.M. Hartmann, A. Gassenq, L. Milord, N. Pauc, V. Reboud, and V. Calvo, J. Cryst. Growth **473**, 20 (2017).

[18] D. Weisshaupt, P. Jahandar, G. Colston, P. Allred, J. Schulze, and M. Myronov, in *2017 40th Int. Conv. Inf. Commun. Technol. Electron. Microelectron.* (IEEE, 2017), pp. 43–47.

[19] J. Aubin, J.M. Hartmann, A. Gassenq, J.L. Rouviere, E. Robin, V. Delaye, D. Cooper, N. Mollard, V. Reboud, and V. Calvo, Semicond. Sci. Technol. **32**, 094006 (2017).

[20] J. Aubin, J.M. Hartmann, J.P. Barnes, J.B. Pin, and M. Bauer, ECS Trans. **75**, 387 (2016).

[21] É. Bouthillier, S. Assali, J. Nicolas, and O. Moutanabbir, Arxiv.Org/Abs/1901.00436 (2019).

[22] J. Aubin and J.M. Hartmann, J. Cryst. Growth **482**, 30 (2018).

[23] W. Wang, Q. Zhou, Y. Dong, E.S. Tok, and Y.-C. Yeo, Appl. Phys. Lett. **106**, 232106 (2015).

[24] S. Biswas, J. Doherty, D. Saladukha, Q. Ramasse, D. Majumdar, M. Upmanyu, A. Singha, T. Ochalski, M.A. Morris, and J.D. Holmes, Nat. Commun. **7**, 11405 (2016).

[25] M.S. Seifner, S. Hernandez, J. Bernardi, A. Romano-Rodriguez, and S. Barth, Chem. Mater. **29**, 9802 (2017).

[26] L. Gagliano, M. Albani, M.A. Verheijen, E.P.A.M. Bakkers, and L. Miglio, Nanotechnology **29**, 315703 (2018).

[27] L. Gagliano, A. Belabbes, M. Albani, S. Assali, M.A. Verheijen, L. Miglio, F. Bechstedt, J.E.M. Haverkort, and E.P.A.M. Bakkers, Nano Lett. **16**, 7930 (2016).

[28] A.C. Meng, C.S. Fenrich, M.R. Braun, J.P. McVittie, A.F. Marshall, J.S. Harris, and P.C. McIntyre, Nano Lett. acs. nanolett.6b03316 (2016).

[29] A.C. Meng, M.R. Braun, Y. Wang, C.S. Fenrich, M. Xue, D.R. Diercks, B.P. Gorman, M.-I. Richard, A.F. Marshall, W. Cai, J.S. Harris, and P.C. McIntyre, Mater. Today Nano (2019).

[30] S. Assali, R. Bergamaschini, M. Albani, M.A. Verheijen, M. Loda, E. Scalise, S. Koelling, E.P.A.M. Bakkers, and L. Miglio, Unpublished (n.d.).

[31] F. Gencarelli, B. Vincent, J. Demeulemeester, A. Vantomme, A. Moussa, A. Franquet, A. Kumar, H. Bender, J. Meersschaut, W. Vandervorst, R. Loo, M. Caymax, K. Temst, and M. Heyns, ECS J. Solid State Sci. Technol. **2**, P134 (2013).

[32] Y. Liang, W.D. Nix, P.B. Griffin, and J.D. Plummer, J. Appl. Phys. **97**, 043519 (2005).

[33] M.J. Süess, R. Geiger, R. a. Minamisawa, G. Schiefler, J. Frigerio, D. Chrastina, G. Isella, R. Spolenak, J. Faist, and H. Sigg, Nat. Photonics **7**, 466 (2013).

[34] J. Petykiewicz, D. Nam, D.S. Sukhdeo, S. Gupta, S. Buckley, A.Y. Piggott, J. Vučković, and K.C. Saraswat, Nano Lett. **16**, 2168 (2016).

[35] K. Guilloy, N. Pauc, A. Gassenq, Y.-M. Niquet, J.-M. Escalante, I. Duchemin, S. Tardif, G. Osvaldo Dias, D. Rouchon, J. Widiez, J.-M. Hartmann, R. Geiger, T. Zabel, H. Sigg, J. Faist, A. Chelnokov, V. Reboud, and V. Calvo, ACS Photonics **3**, 1907 (2016).

[36] N. von den Driesch, D. Stange, S. Wirths, D. Rainko, I. Povstugar, A. Savenko, U. Breuer, R. Geiger, H. Sigg, Z. Ikonic, J.-M. Hartmann, D. Grützmacher, S. Mantl, and D. Buca, Small **13**, 1603321 (2017).

[37] A. Attiaoui and O. Moutanabbir, J. Appl. Phys. **116**, 063712 (2014).

# Supporting information:

# Strain engineering in Ge/GeSn core/shell nanowires

S. Assali,[1,2,¥,*] M. Albani,[3,¥,*] R. Bergamaschini,[3] M. A. Verheijen,[1,4] A. Li,[1,5,6] S. Kölling,[1] L. Gagliano,[1] E. P.A.M. Bakkers,[1,6] and L. Miglio[3]

[1] Department of Applied Physics, Eindhoven University of Technology, 5600 MB Eindhoven, The Netherlands
[2] Department of Engineering Physics, École Polytechnique de Montréal, C. P. 6079, Succ. Centre-Ville, Montréal, Québec H3C 3A7, Canada
[3] L-NESS and Dept. of Materials Science, University of Milano Bicocca, 20125, Milano, Italy
[4] Eurofins Materials Science BV, High Tech Campus 11, 5656AE Eindhoven, The Netherlands
[5] Beijing University of Technology, Pingleyuan 100, 100124, P. R. China
[6] Kavli Institute of Nanoscience, Delft University of Technology, 2600 GA Delft, The Netherlands

## Contents

## S1. Experimental methods

**Growth.** Samples are grown using an Aixtron CCS-CVD reactor with $H_2$ as a carrier gas. 1% $GeH_4$, $SnCl_4$, and 1% HCl are used as precursor gases. The nucleation of the (tapered) Ge NW base is held at 425°C and at a reactor pressure of 100 mbar using $GeH_4$ precursor. Next, the sample is cooldown to 320°C for the 2 hours growth of the untampered Ge NWs, using a reactor pressure of 75 mbar. The GeSn shell is grown at 300°C and at a reactor pressure of 50 mbar using $GeH_4$, $SnCl_4$, and HCl precursors.[1] A Ge/Sn ratio in gas phase of 740 for 5 hours was used for the samples in Figs. 1-2, while the sample in Fig. 3 was grown with Ge/Sn=450 for 2 hours.[2]

**TEM characterization.** For TEM measurements a probe-corrected JEOL ARM 200F Transmission electron microscope equipped with a 100 $mm^2$ SSD Centurio Energy Dispersive X-ray Spectroscopy (EDS) detector was used (operated at 200 kV). Cross-sectional TEM lamella samples were prepared using a dual-beam focused ion beam (FIB) apparatus (FEI Nova Nanolab 600i). The lamella was cut-out at an operating voltage of 30 kV, during the thinning of the TEM window the voltage was successively lowered to 5kV. A 500 nm thick protective layer was deposited on the sample surface before FIB milling using first electron beam and then ion beam induced metal deposition.

## S2. Ge core NWs with different diameters.

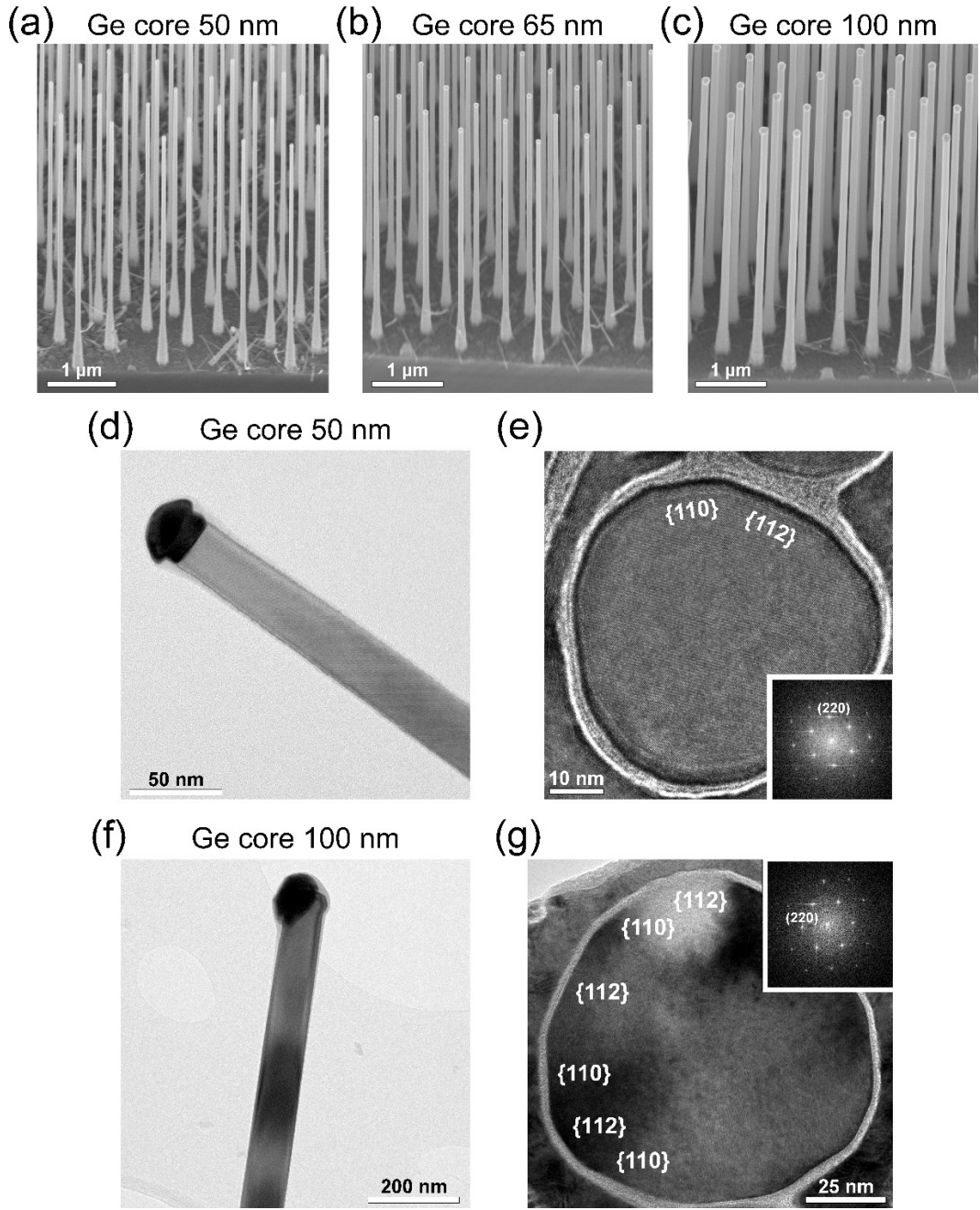

Figure S1. (a-c) SEM images of the Ge NW arrays (tilting angle 30°) grown in the nanoimprint pattern showing flat, vertical sidewalls at diameters of 50 nm (a), 65 nm (b), and 100 nm (c). (d) Low-resolution TEM image acquired along the [110] zone axis of a Ge NW with a 50 nm diameter. (e) Cross-sectional HRTEM image and corresponding FFT pattern (inset) of a Ge NW with a 50 nm diameter. (f) Low-resolution TEM image acquired along the [110] zone axis of a Ge NW with a 100 nm diameter. (g) Cross-sectional HRTEM image and corresponding FFT pattern (inset) of a Ge NW with a 100 nm diameter.

## S3. Effect of HCl on the GeSn shell growth.

The effect of HCl on the shell growth was investigated on a sample grown for 2 h using a Ge/Sn ratio of 1285, as shown in Fig. S2. When no HCl is supplied in the gas phase (Fig. S2a) a small increase of the NW diameter is observed in the proximity of the Au-Sn droplet compared to the regular growth protocol using HCl (Fig. S2b). This is most likely induced by the catalytic effect[3] of the Au-Sn droplet that enhances the precursors decomposition, hence it (locally) increases the shell growth rate. However, no significant change in the morphology of the bottom section of the NWs is observed, thus excluding a possible etching effect of HCl[4] on GeSn.

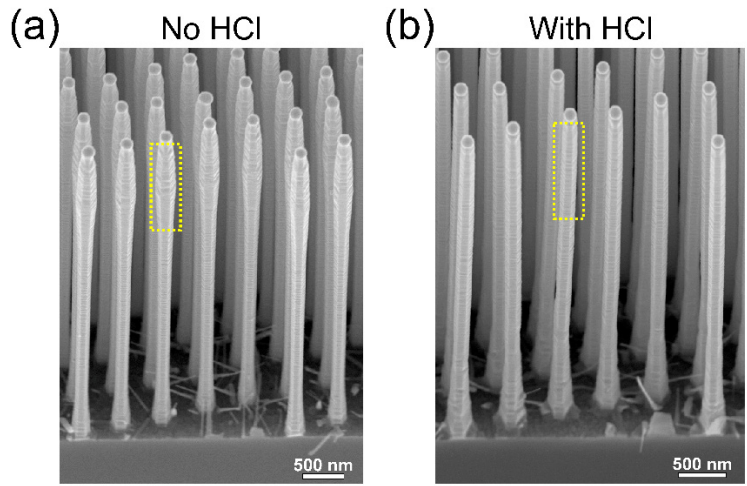

Figure S2. (a-b) SEM images for two different samples grown without (a) and with (b) HCl as a precursor gas.

## S4. Finite Element Method simulations

The strain relaxation in the core/shell NW structure is modelled numerically by finite element method, exploiting COMSOL Multiphysics® software. The structure of the NW is made of a faceted core, surrounded by a shell with six {112} and six {110} facets, all perpendicular to the (111) substrate. The structure is simulated as fixed to the substrate by Dirichlet boundary conditions and the NW is 4 µm long. An adaptive 3D mesh is used to ensure the convergence of the numerical strain results. To computer the deformations in the structure, a mechanical equilibrium problem is solved, by considering a zero-stress condition on the free surfaces of the NW. The strain is caused by the lattice parameter difference between the pure Ge in the core and the GeSn alloy in the shell. To compute the latter, we have used a corrected version of the Vegard's law, as a function of the Sn composition x, $a_{GeSn} = a_{Ge}(1 - x) + a_{Sn}x + b_{GeSn}\, x(1 - x)$ with $a_{Ge} = 5.657$ Å and $a_{Sn} = 6.489$ Å the bulk lattice parameters and $b_{GeSn} = 0.041$ Å the bowing parameter.[5] The lattice mismatch is set in the simulations as an initial condition, corresponding to a compressive strain in the shell. Then the numerical procedure determines the equilibrium deformation that minimizes the elastic energy. To this goal, we have also set the elastic constants for the cubic GeSn alloy as a linear interpolation with composition between the bulk values of Ge and Sn ($C_{Sn}^{11} = 69$ GPa, $C_{Sn}^{12} = 29.3$ GPa, $C_{Sn}^{44} = 36.2$ GPa, $C_{Ge}^{11} = 126$ GPa, $C_{Ge}^{12} = 44$ GPa, $C_{Ge}^{44} = 67.7$ GPa)[6]. The components of the strain tensor are expressed as radial, tangential and axial by considering the 12-fold symmetry of the NW cross-section. For each facet the tangential and axial components give a similar quantification as the in-plane strain for planar film, while the radial direction exhibits a tensile strain as a consequence of the compression in the other directions, due the Poisson's ratio and profiting of the free surface.

# References.

[1] S. Assali, A. Dijkstra, A. Li, S. Koelling, M.A. Verheijen, L. Gagliano, N. von den Driesch, D. Buca, P.M. Koenraad, J.E.M. Haverkort, and E.P.A.M. Bakkers, Nano Lett. **17**, 1538 (2017).

[2] S. Assali, R. Bergamaschini, M. Albani, M.A. Verheijen, M. Loda, E. Scalise, S. Koelling, E.P.A.M. Bakkers, and L. Miglio, Unpublished (n.d.).

[3] M.A. Verheijen, G. Immink, T. De Smet, M.T. Borgstrom, and E.P.A.M. Bakkers, J. Am. Chem. Soc. **128**, 1353 (2006).

[4] M.T. Borgström, J. Wallentin, J. Trägardh, P. Ramvall, M. Ek, L.R. Wallenberg, L. Samuelson, and K. Deppert, Nano Res. **3**, 264 (2010).

[5] F. Gencarelli, B. Vincent, J. Demeulemeester, A. Vantomme, A. Moussa, A. Franquet, A. Kumar, H. Bender, J. Meersschaut, W. Vandervorst, R. Loo, M. Caymax, K. Temst, and M. Heyns, ECS J. Solid State Sci. Technol. **2**, P134 (2013).

[6] O. Madelung, *Semiconductors* (Springer Berlin Heidelberg, Berlin, Heidelberg, Heidelberg, 1991).